# A Novel 4-D Dataset Paradigm for Studying Complete Ligand-Protein Dissociation Dynamics

Maodong Li [1, †], Jiying Zhang [2, †], Zhe Wang[1,3, †], Bin Feng [2], Wenqi Zeng [2], Dechin Chen[1], Zhijun Pan[1], Yu Li[2,*], Zijing Liu[2,*], Yi Isaac Yang [1,*]

[1] Institute of Systems and Physical Biology, Shenzhen Bay Laboratory, Shenzhen 518132, China.

[2] International Digital Economy Academy, Shenzhen 518000, China.

[3] Hong Kong University of Science and Technology, Hong Kong 000000, China

†These authors contributed equally: Maodong Li, Jiying Zhang

* Corresponding author. Yi Isaac Yang: yangyi@szbl.ac.cn, Zijing Liu: liuzijing@idea.edu.cn, Yu Li: liyu@idea.edu.cn

**Author Contributions:** Y.I.Y., Z.L., Y.L., M.L., and J.Z. conceived the work. M.L. setup the dataset process. J.Z. and M.L. performed all molecular simulations on the GPU cluster of IDEA company. Y.I.Y., Z.L., M.L., and J.Z. analyzed the dataset. M.L., and D.C. carried out the NEB application to get average pathways. Z.L., J.Z., B.F., W.Z., and M.L. carried out the AI generative model. M.L., Y.I.Y., D.C., and Z.P. contributed the module of the software SPONGE. Y.I.Y., Z.L., M.L., J.Z., Z.W., Y.L., B.F., W.Z., D.C. and Z.P. interpreted the results. Y.I.Y., M.L., J.Z., Z.L. and Z.W. wrote the paper.

**Competing Interest Statement:** The authors declare no competing interests.

**Keywords:** AI-Normalized dataset, dissociation kinetics, drug design, generative model

## Abstract

The kinetics and dynamics of drug-protein binding and dissociation are crucial to understanding drug absorption and metabolism. Despite advances in artificial intelligence (AI) tools for drug-protein interaction studies, existing training datasets remain limited to static structures or quasi-static conformations. This paper proposes a novel computational approach for rapidly generating drug-protein dissociation trajectories and presents the inaugural dynamically time-resolved 4-D (*t, x, y, z*) trajectory database DD-13M. This dataset captures over 26,000 complete dissociation processes for 565 ligand-protein complexes, providing nearly 13 million frames of all-atom simulation trajectories. A deep equivariant generative model, UnbindingFlow, was trained using the DD-13M dataset. This model has the capacity to produce dissociation trajectories for novel targets whilst accurately predicting their rate constants ($k_{off}$). DD-13M introduces a new type of training dataset for AI models, establishing a *de novo* paradigm for studying the dynamics of drug-protein interactions.

## Introduction

Research into ligand-protein binding (LPB) interactions constitutes a foundation in modern drug discovery, with computational models serving as indispensable research tools in this filed. Typically, there are two main goals of computational studies dealing with ligand-protein interactions: the identification of the ligand-protein binding modes and the calculation of the thermodynamics and kinetics of their binding/dissociation processes. Binding mode refers to the binding site of the ligand molecule within the protein pocket and the corresponding optimal binding conformation. Currently developed technologies cover the entire process from lead compound optimization to binding affinity refinement. In particular, molecular docking has become a crucial step to generate potential drug candidates for lead compounds, for example, Autodock(1), Glide(2), DSDP(3). In recent years, artificial intelligence (AI) has revolutionized structural biology (4) (5) and drug discovery (6-10). The success in single biomolecular structures prediction has shifted the recent research focus toward understanding the complex, including ligand-protein binding. The computational speed of these binding mode prediction tools is typically rapid, making them widely used in virtual drug screening. However, such tools generally only provide low precision “binding affinity” and cannot offer the accurate values of binding free energy, which directly determines the strength of ligand binding within the protein pocket.

Therefore, the precise calculation of the thermodynamics of ligand-protein interactions, primarily the binding free energy, is the traditional core task in drug discovery following the identification of the binding mode. The computational methods for binding free energy that have been developed in the past 30 years can be divided into two main categories, namely, alchemical methods and physical pathway methods. Alchemical methods are based on thermodynamic cycles, so they typically employed to quantify the relative binding free energy of a target ligand against a reference ligand binding to a protein. For example, free energy perturbation (FEP) (10, 11) and thermodynamic integration (TI)(12) methodologies have been extensively employed for the optimization of drug molecular structure. In contrast, the physical pathway method aims to estimate binding/dissociation free energies by reproducing the minimum free energy pathway (MFEP) of the ligand toward the protein target, enabling atomistic resolution of transient intermediate states and mechanistic discrimination among multiple pathways(13-18). Most of these methods involve molecular dynamics (MD) simulations with specific enhanced sampling techniques, such as example, umbrella sampling, steered molecular dynamics (19-22), PathCV-based metadynamics (23), and funnel metadynamics (24-27).

However, biological processes are inherently dynamic. The kinetics and dynamics of ligand-protein binding and dissociation processes govern the absorption and metabolic processes following drug administration, which can influence the therapeutic efficacy. This holds significant physiological importance in advancing rational drug design and related fields. Unfortunately, the kinetics and dynamics studies related to ligand-protein interactions are markedly fewer compared to thermodynamics. Experimental data on ligand dissociation rate constants ($k_{\text{off}}$) are far less abundant than those on binding affinity(28, 29). In terms of computational methods, there are also significant challenges in accurately calculating the kinetics and dynamics of binding or dissociation. Early algorithms based on spatial coordinates and scoring functions can identify dissociation pathways with relatively low computational costs. For instance, GPathFinder(30) employs geometric space search algorithms to enumerate possible dissociation pathways for a given drug-protein complex. However, such methods can only provide thermodynamic local optima on candidate pathways and fail to resolve issues related to the kinetic continuity of dissociation pathways. In contrast, MD simulations can capture the true kinetic interactions between drugs and proteins (31), making them the most intuitively persuasive approaches for accurately.

Although MD simulations represent the one of the most competitive methods for studying the kinetics and even dynamics of ligand-protein interactions, their immense computational cost presents a bottleneck, rendering them impractical for the high-throughput screening required in drug discovery(32, 33). Recently, generative AI models have emerged. If generative AI models can learn the underlying physical principles from a large-scale data of MD simulations, it may be possible to achieve efficient prediction the dynamic processes of binding/dissociation. Current attempts include DynamicBind(34), Bioemu(35), ProtMD(36), NeuralMD(37), etc.

However, there is currently a lack of suitable training datasets for training a generative AI model capable of predicting the complete dissociation process of ligands from binding pockets. Among the earliest standardized datasets, PDBbind+(38, 39) has emerged as a cornerstone for static docking score benchmarking(40-42). Building upon this foundation, MISATO(43) introduced trajectory data through localized conformational relaxation of selected structures from PDBbind dataset. The MISATO dataset has enabled novel applications, particularly in generative models such as NeuralMD(37), which leverages trajectory noise for learning and has garnered considerable attention. Several datasets have recently been developed to push the boundaries of molecular simulation: the DynaRepo dataset(44) expands each simulated trajectories to a length of 500ns; the PLAS-20k dataset(45) employs MMPBSA for improved the accuracy of energy labels; and the Navigating Protein Landscapes dataset(46) attempts to explore protein allosteric space through multi-scale coarse-grained models. However, these databases share an inherent limitation: their primary metric for assessing simulation validity relies on RMSD, which inherently constrains conformational sampling to minimal deviations from the initial state (L-P). Consequently, these trajectories(43, 44, 46) are better characterized as "quasi-static" relaxation and fail to completely capture the dynamic dissociation process (L-P → L + P). This creates a critical gap: there is no large-scale, public dataset of *complete* unbinding trajectories to train the next generation of predictive AI models.

To bridge this critical gap, we propose a novel “AI+physics” research paradigm for studying ligand-protein dissociation kinetics and dynamics. (Figure 1) We first implemented an enhanced sampling strategy based on MetaD (metadynamics) (47) method to rapidly generate dissociation trajectories for diverse drug-protein complexes, thereby establishing an efficient automatic MD simulation pipeline. Utilizing this pipeline, we present **DD-13M**, the first large-scale, dynamically time-resolved 4D ($t$, $x$, $y$, $z$) dataset capturing complete ligand-protein dissociation dynamics. We further propose a novel binding pocket angiography technique to derive 3D topological landscapes of binding affinities through, thereby enabling statistically robust clustering of ligand escape pathways within this dataset. To showcase the dataset's utility, we developed **UnbindingFlow**, a novel deep equivariant generative model trained on DD-13M, capable of generating physically plausible, collision-free dissociation trajectories. This work provides the foundational data, comprehensive analysis tools and an accurate model to move computational drug discovery beyond static representations and into the crucial domain of molecular dynamics.

## Results

### MD Simulation Pipeline for Rapid Generation of Drug Protein Dissociation Trajectories

To rapidly generate dissociation trajectories for drug-protein complexes, we design an automatic MD simulation pipeline based on enhanced sampling. Constrained by current computing hardware, conventional MD simulations struggle to achieve the complete dissociation process of even the simplest complex systems within an acceptable computational timeframe. Here, we propose a simple enhanced sampling strategy based on the well-established MetaD method, enabling efficient dissociation of small molecules from binding pockets. We directly use the Cartesian coordinates $\boldsymbol{R}_{\mathrm{c}} = (x, y, z)$ of the center of mass (COM) of the small ligand molecules, a 3D variable, as the collective variables (CVs) (48) for MetaD(47, 49). Therefore, performing MD simulations with the regular MetaD results in the continuous accumulation of three-dimensional Gaussian-type repulsive potentials $\{G(\boldsymbol{R}_c; t)\}$ in the 3D CV space of $\boldsymbol{R}_c$, that is, forming an ever-expanding bias potential $V(\boldsymbol{R}_c; t)$ which rapidly "pushes" the ligand molecule out of the binding pocket:

$$V(\boldsymbol{R}_c; t) = \sum_t G(\boldsymbol{R}_c; t) = \sum_t w e^{-\frac{1}{2}\left\| \frac{\boldsymbol{R}_c - \boldsymbol{R}'_c(t)}{\boldsymbol{\sigma}} \right\|^2} \qquad [1]$$

where $\boldsymbol{R}'_c(t)$ is the value of the CVs $\boldsymbol{R}_c$ at the simulation step $t$, and $w$ as well as $\boldsymbol{\sigma}$ is the weight coefficient and standard deviation of the Gaussian function $G(\boldsymbol{R}_c; t)$, respectively. It should be noted that a more popular variant, well-tempered metadynamics (WT-MetaD) (50), uses a time-dependent coefficient $\omega(t)$ as the Gaussian weights, but here we still use a fixing weight coefficient $w$. This will not only accelerate the ligand's escape from the protein pocket but also provide a way to estimate the 3D binding free energy landscape of the ligand within the protein pocket, which will be detailed later.

This approach is a simple but highly effective enhanced sampling strategy for achieving the complete dissociation process of ligand-protein complexes. Typically, using MetaD to sample three-dimensional CVs is inefficient. However, as the space within protein pockets is usually very small, employing the COM $\boldsymbol{R}_c$ as the CV for MetaD and performing MD simulations starting from the binding conformation can quickly eject the ligand from the protein pocket (Fig. 2A). Of course, one single simulation can produce only a random dissociation trajectory. However, by subjecting the initial positions and velocities of the ligands to random perturbations, it is possible to obtain a range of dissociation trajectories for the ligand-protein complex. The setup for this effective ligand-protein dissociation strategy is simple and routine, applying to almost all enhanced sampling software or library, such as PLUMED(51, 52) or COLVARS(53).

Based on the sampling strategy, we established an automatic MD simulation pipeline using the molecular modelling software SPONGE(54) to efficiently generate ligand-protein dissociation trajectories(Fig. 2B). SPONGE is a next-generation molecular modelling program specifically optimized for artificial intelligence and enhanced sampling, enabling efficient execution of the MetaD method. We have constructed an automatic MD simulation pipeline based on SPONGE, which can automatically model ligand-protein complex systems and then perform MD simulations. It will automatically terminate the simulation once it determines the ligand has escaped the protein pocket. For details regarding the modeling, running and stopping of this automatic MD simulation pipeline, please refer to the Supplementary Information Simulation details Section.

This pipeline enables high-throughput MD simulations, which can produce a series of trajectories for different dissociation pathways within the same ligand-protein complex system. When the pipeline models a complex prior to one MD simulation, it will make a tiny random perturbation to the initial positions of the atoms in the bound ligand molecule and assign a random velocity to all atoms in the simulation system based on a Boltzmann distribution. In this way, each MD simulation will yield a different dissociation trajectory. Therefore, a multitude of MD simulation trajectories can be obtained, encompassing diverse dissociation processes of the ligand-protein complex. These diverse MD simulation trajectories can be employed to analyze distinct dissociation pathways of ligand-protein interactions, as well as to investigate the thermodynamics and kinetics of the binding/dissociation process.

**DD-13M: A 4-D Ligand-Protein Dissociation Dataset**

By employing the automatic high-throughput MD simulation pipeline featuring our enhanced sampling strategy, we generate a public dataset of drug-protein dissociation dynamics. As we aim to construct a dataset that can be analyzed for dynamics, our dataset is based on the $k_{off}$ subset(28) of the PDBbind(38) dataset, using the 680 ligand-protein 3D structures as the initial conformation for MD simulations. The PDBbind dataset provided the initial docking conformation by experimental structures and validated through MD simulations. We further select the $k_{off}$ subset because it contains experimental dissociation kinetic constants, which thermodynamically implicate the existence of substantial energy barriers in these systems. Even for the fastest system ($k_{off} = 8.9 \times 10^2$ $s^{-1}$, 5M04_405) in the $k_{off}$ databases, a typical unbinding event occurs on microsecond timescales, well beyond the reach of conventional MD (requiring years on regular computer). In contrast, our enhanced sampling algorithm achieves this feat with an average time of just 45 minutes per dissociation trajectory, achieving hundreds of thousands of times acceleration relative to conventional MD, see Fig. 1A.

Our DD-13M database offers three key advantages for studying ligand-protein dissociation: 1. it ensures the integrity of the complete escape pathway from the bound to the fully dissociated state; 2. by employing enhanced-sampling algorithm, it captures long-timescale dissociation processes within a short simulation window; 3. it significantly improves the sampling probability of rare transition-state events through the embedding of an escape boundary criterion. The coordinates were collected every 100 frames (0.1ps). For each complex, we performed 50 paralleled MetaD runs with different random seeds.

The dissociation dynamics dataset DD-13M is a dedicated trajectory database focused on the drug-protein dissociation process. Each of the 680 complexes from the $k_{off}$ dataset underwent 50 parallel replica simulations for one-time dissociation. Our workflow successfully modelled 95.4% (649 out of 680) of the complexes in the database. A total of 26,612 dissociation trajectories were collected, yielding 12,786,863 frames of complex conformations, shown in Fig. 3. The dissociation dynamics dataset DD-13M is publicly available at https://huggingface.co/datasets/SZBL-IDEA/MD (For facilitated browsing and categorical download, a dedicated web interface is maintained at: https://aimm.szbl.ac.cn/database/ddd/#/home).

Analysis of the DD-13M dataset revealed a median trajectory length of 21.8 ps, with 94 trajectories (0.35%) exceeding 1.0 ns. The average number of effective trajectories per complex was 47.04. The average clash score(8) (see the Method section for algorithmic details, with higher values indicating more pronounced van der Waals(55) clashes between the ligand and the protein(56, 57)) for most trajectories was approximately $0.336 \pm 0.045$ . This indicates that our MD simulation successfully generates pathways with small geometric clashes. Even if the enhanced sampling technique is used to generate energy disturbances, the trajectory conformation still maintains a small atomic collision and can be maintained within the range of kinetic accessibility.

## Binding Pocket Angiography: Revealing 3D Binding Affinity Landscapes

The DD-13M dataset generated from MD simulations contains wealthy thermodynamic and kinetic information about drug-ligand binding/dissociation. Theoretically, the binding affinity $k_b$ is positively proportional to the concentration ratio of the drug being outside versus inside the pocket, $k_b = \frac{[LP]}{[L][P]}$. However, for faster computations, conventional methods such as docking(3, 58), FEP(10, 11), and TI(12)often estimate binding affinity based solely on a single, optimal docking pose, effectively reducing the complex. However, the local binding energy varies significantly at different locations within the pocket, a critical detail for understanding dissociation dynamics. Furthermore, the accurate calculation of absolute binding free energy should be represented by the statistical sum of probabilities across all relevant positions within the binding pocket. DD-13M incorporates MD simulation trajectories produced via the enhanced sampling method metadynamics (MetaD), providing the complete 3D free energy landscape of the small molecule within the pocket.

Once we obtain sufficient 4-D (*t, x, y, z*) escape trajectory replica, the 3-D free energy surface (FES) for the conformational distribution of the ligand binding to the protein pocket can be calculated. As we employ the conventional MetaD method with fixed Gaussian weight $w$, the FES $F(\boldsymbol{R}_c)$ should be proportional to the negative of the bias potential $V(\boldsymbol{R}_c)$ when the simulation time approaches infinity: $F(\boldsymbol{R}_c) \propto -\lim_{t\to\infty} V(\boldsymbol{R}_c; t)$. Evidently, our dissociation trajectories are all too short to satisfy this condition, as the simulations cease once the ligand departs the protein pocket. However, if there are enough stochastic simulation trajectories, the bias potential $\lim_{t\to\infty} V(\boldsymbol{R}_c; t)$ from long-term simulations should be proportional to the average value $\langle V(\boldsymbol{R}_c)\rangle$ of the set of bias potentials $\{V_i(\boldsymbol{R}_c)\}$ corresponding to the short random dissociation trajectories. Therefore, the 3D binding FES $F(\boldsymbol{R}_c)$ can be estimated as:

$$F(\boldsymbol{R}_c) \propto -\langle V(\boldsymbol{R}_c)\rangle \approx -\lim_{N\to\infty} \frac{1}{N}\sum_{i}^{N} V_i(\boldsymbol{R}_c) \qquad [2]$$

This approach of obtaining 3D binding free energy landscapes by filling the Gaussian-type repulsive potentials is similar to the angiography techniques in clinical examinations, hence we named it “binding pocket angiography” (BPA). shown as Fig. 4A. This approach establishes a novel free energy computational strategy for MetaD: when individual sampling replicas achieve sufficiently rapid execution, the collective probability P from multiple replicas can efficiently approximate the free energy surface F. Therefore, BPA also offers an effective solution to the convergence difficulties encountered in traditional MetaD methods.

BPA provides a detailed 3D landscape of binding affinity for studying drug-protein interactions. This 3D binding affinity landscape can be analogized to comprehensive navigational atlases of drug-protein binding complex systems, enabling further analyzing their thermodynamics and kinetics of this binding system akin to AI-powered autonomous driving systems. For example, we can calculate the minimum free energy paths (MFEPs) for binding/dissociation using path-searching methods like nudged elastic band (NEB)(59, 60) and string method(61, 62). A detailed analysis of the ligand-protein dissociation pathway is provided in the following subsection.

In addition, the BPA corroborates the “ligand cloud” concept(63), indicating that ligand-protein binding is a dynamic process requiring ensemble description rather than a static lock-and-key model. Moreover, this technology further provides more comprehensive 3-D free energy contour information. This spans from the ligand cloud's binding hot spot to the protein's outer surface, becoming quantitatively accurate in the physical sense as the replica count $N$ approaches infinity, see Fig. S1.

**Protein-Drug Dissociation Pathways**

To further explore pathway pocket via BPA, we refined 478 robust dissociation pathways from over 26,000 MD simulation trajectories, shown in Fig. 4 and Fig. S3. This infrastructure pioneers a new paradigm for drug-protein trajectory datasets, transforming raw simulation data into quantitatively verifiable results for completely dissociation kinetics.

We projected the endpoints of 26,612 trajectories from 565 complexes onto a predefined surface and performed clustering to identify distinct dissociation pathways. Among these, 493 complexes had trajectories with Mean Squared Error (MSE) <200, enabling direct averaging of their parallel trajectories. For the remaining 71 complexes, their escape paths featured more than two exits (multiple-pathway systems), and clustering was performed based on exit positions. This processing yielded a total of 763 exit pathways across the 565 complexes.

Subsequently, we applied the Nudged Elastic Band (NEB) method(59) to obtain these 763 MFEPs, as well dissociation pathways, shown in Fig.3I. Notably, 221 of these NEB-averaged paths had lengths shorter than 5.0 Å, indicating that nearly half of complex systems exhibit no dominant pathways, shown as Fig.4E, It reveals that the ligands in these complexes were predominantly located at the protein's outer surface and their properties are primarily determined by the binding affinity strength and the diffusion rate of ligands in solvents, tagged as "shallow pocket" in Fig.3C and "short pathway" in Fig.3E. For such proteins, the dominant factor influencing their dissociation process was diffusion kinetics rather than pathway selection. In contrast, there are less than 5% of complex with very large pockets given that $\Delta F > 0$ at the binding position, the conventional path energy barrier assumption is not suitable for calculations, shown in Fig.4F. To ensure dynamic reproducibility, we excluded clusters with only one trajectory. After excluding these short or single-visit pathways, our final dissociation pathway subset consisted of 478 trajectories from 338 drug-protein complexes. This statistical result aligns with our expectations from the $k_{off}$ dataset, indicating a significant population of deeper drug-protein binding pockets suitable for constructing ligand dissociation pathway dataset. The median path length within this subset was observed to be 11.98 Å, with the longest path reaching 32.69 Å. Among the 338 complexes, 270 exhibited unique dissociation pathway dominance, while 68 demonstrated reproducible multi-pathway dissociation features, including an extreme case in the 6f7b system with a notable cluster of 7 distinct pathways, see Fig.S3.

Then, using mass centroid constraints, we performed local conformational relaxations at each step to generate plausible intermediate structures along the dissociation pathway. This approach allowed us to construct continuous, collision-free all-atom trajectories that accurately represent ligand dissociation dynamics. These trajectories were stored in the h5md format. In Fig. 4, we present all dissociation pathways for a specific drug-protein complex (6OOY). Among 50 successful dissociation trajectories, three major clusters were identified, see Fig.4B-D and Fig. S2. By design, the DD-13M database prioritizes system universality, and thus the provided energy barriers are intended for qualitative comparison. Quantitative ensemble calculations and kinetics derivations are an immediate priority for future development.

## UnbindingFlow: AI Generative Model for Predicting Ligand Dissociation Trajectories

Although our pipeline can generate complete trajectories of small-molecule ligands dissociating from protein pockets at speeds far exceeding conventional MD simulations, its computational time is still challenging to satisfy the demands of drug discovery and design. Here, we developed UnbindingFlow—a deep equivariant generative model that learns the temporal evolution of ligand dissociation directly from MD simulation trajectory. The schematic diagram illustrating the working principle of UnbindingFlow is shown in Fig. 5. The protein–ligand complex is represented in a coarse-grained manner: protein side chains are parameterized in the torsion angle space with a fixed backbone, while the ligand is modeled as a rigid body in the SE(3) space with additional torsional flexibility. These features are processed by an SE(3)-equivariant encoder that maps each frame into a symmetry-preserving latent representation. Rather than learning a noise-to-structure mapping between endpoint conformations, UnbindingFlow is trained on frame pairs sampled from MD simulation trajectories to regress the displacement vector field between them. Following the coarse-grained decomposition introduced in DynamicBind(34), this vector field comprises ligand translation ($\mathcal{L}_{tr}^{\ell}$), rotation ($\mathcal{L}_{rot}^{\ell}$), torsion ($\mathcal{L}_{tor}^{\ell}$), and protein side-chain chi angles ($\mathcal{L}_{chis}^{p}$), thereby directly learning the local dynamics that govern the dissociation process. To capture the directionality and path-dependence of ligand motion, a history aggregation module incorporates a sliding window of preceding ligand positions into the current prediction via attention (Fig. 5, Train), providing temporal context that is inaccessible from any single snapshot. During inference (Fig. 5, Generate), starting from the bound state, the model autoregressively applies the learned vector field and advances the system through continuous ODE integration, updating the ligand pose and protein side chains at each step until complete dissociation.

We trained UnbindingFlow using the MD simulation trajectories from the DD-13M dataset generated by our program pipeline as the training set. After training, the model can generate full trajectories in less than 5 minutes on a single GPU, whereas comparable MD simulations require above 30 minutes. We evaluated whether UnbindingFlow can discover novel exit pathways rather than merely memorizing training data. For the 3wze complex, where training data clustered into three distinct exit paths, UnbindingFlow successfully reproduced these known pathways and, importantly, generated a physically plausible novel trajectory distinct from all existing clusters (Fig. 6). This demonstrates that the model has learned the underlying physics of unbinding rather than overfitting to specific routes. We further validated the model on complexes outside the DD-13M dataset, including 3uod (Fig. 6), 2ao6, and 7abp (Fig. S4). Across all cases, UnbindingFlow consistently generated plausible dissociation pathways with low steric clashes (Clash Score < 0.5 for 95% of frames).

## UnbindingFlow as a Pre-trained Model to Predict the Dissociation Rate Constant $k_{\mathrm{off}}$

Unlike the binding affinity, which correlates with the equilibrium ensemble at the binding pocket, dissociation kinetics depend on the complete dynamic process of the ligand departing from the protein pocket. A more substantial obstacle is posed by the scarcity of reported experimental data, which hinders the accurate prediction of the dissociation rate constant ($k_{\mathrm{off}}$) for ligand-protein interactions. This has been a formidable challenge for traditional computational methods and predictive models. Since the DD-13M dataset incorporates the MD simulation trajectory of the entire process of ligand escape from the binding pocket, the UnbindingFlow model trained on DD-13M capture the latent representations that encode rich dissociation dynamical priors—including transition-state geometries and the kinetic landscape along the unbinding pathway. Therefore, UnbindingFlow serves as a pre-trained model for predicting the dissociation rate constant $k_{\text{off}}$.

We exploit the prediction of $k_{\text{off}}$ values by appending a regression head to the pre-trained encoder (Fig. 5, top): the model accepts a static crystal structure as input, and the encoder extracts representations enriched with the dynamical priors learned from DD-13M, from which the regression head directly predicts the scalar $k_{\text{off}}$ value. Following standard evaluation protocols, our fully fine-tuned model ("UF+Finetune") achieves a Pearson correlation ($R_p$) of 0.826 on the validation set (Table 1, Fig. 7), significantly surpassing the baseline published by Liu *et al.* (28)($R_p$ = 0.524). Furthermore, the model also improved performance and generalized to the challenging HIV-1 test set ($R_p$ = 0.339 vs. reported 0.264). The performance gain arises because DD-13M captures the complete spatiotemporal evolution of ligand dissociation, including high-energy transition states that are rarely sampled by conventional methods, enabling the model to learn the full kinetic landscape from which rates can be inferred even from static inputs.

To validate the critical role of dissociation trajectories in predicting $k_{\text{off}}$ values, we conducted a progressive ablation experiment to rigorously isolate the contribution of DD-13M (Table 1). We trained a model "UF w/o Pretrain" from scratch using only 332 static structures and $k_{\text{off}}$ labels from the PDBbind-koff-2020(28) dataset. This model shares the same static structures as UnbindingFlow but is not pretrained using the DD-13M dataset. "UF w/o Pretrain" model yielded a validation $R_p$ of only 0.256, confirming that the architecture alone without dynamical priors struggles to make accurate regression predictions on such a small dataset. In contrast, after pretraining with MD trajectory simulations from DD-13M, the Pearson correlation coefficient of the model ("UF+Finetune") increased to $R_p$ =0.826. Remarkably, even freezing all pre-trained weights and optimizing only the regression head, the model ("UF+Linear") still achieved a Pearson correlation coefficient $R_p$ = 0.670, surpassing existing baselines. This demonstrates that the representations learned from DD-13M inherently encode kinetic information without any task-specific parameter updates. This hierarchy—where even frozen features outperform baselines and fine-tuning yields state-of-the-art (SOTA) results—provides strong evidence that the 4D dynamical information in DD-13M is indispensable for capturing the physics of ligand unbinding.

## Discussion

In this study, we introduce a new paradigm for investigating ligand-protein dissociation by combining a novel high-throughput simulation strategy with deep generative modeling. We first proposed an efficient enhanced sampling method for achieving drug-protein dissociation and utilized it to establish a MD simulation pipeline that rapidly generates dissociation trajectories. Employing this MD simulation pipeline, we present DD-13M, the first large-scale open-source dataset dedicated to ligand-protein dissociation dynamics, containing 26,612 complete trajectories across 565 complexes. Based on DD-13M, we developed UnbindingFlow, a novel a deep equivariant generative model, to generate diverse and collision-free dissociation trajectories. Our work addresses a critical gap in computational drug discovery, moving beyond static or quasi-static representations to the continuous, dynamic process of a ligand unbinding from its target protein.

The DD-13M dataset is generated by a multi-replica Metadynamics approach, which offers significant practical advantages over conventional, single-run enhanced sampling methods, including accelerated convergence and resource-efficient execution on accessible hardware. The multi-replica MetaD ensembles achieve accelerated convergence compared to serial time-extension strategies(15, 50, 64) by decomposing long sequential simulations into discrete, non-communicating trajectories, which are distributed across a single GPU to substantially reduce the time-to-solution. From this dataset, we also introduce "binding pocket angiography" (BPA), a novel method to visualize the affinity pocket of the protein as a 3D free energy contour, providing comprehensive information on the dissociation landscape. Then we apply the NEB method to obtain these 478 MFEP pathways. It should be noted that although this paper employs shooting approach using the conventional MetaD for angiography of binding pockets, the acquisition of BPA can be achieved through various sampling methods. For instance, if we are presently planning to seal the protein's outer surface and perform back-and-forth convergence sampling using the sinkMeta method(65), then we can derive a higher-precision BPA.

Conventional datasets derived from experimental sources often require extensive cleaning and normalization to ensure consistency, a process that can introduce bias and limit scalability. In contrast, our DD-13M dataset, sourced from rigorously controlled molecular simulations, is inherently self-consistent and requires no such preprocessing, making it immediately suitable for deep learning. Leveraging this advantage, we trained UnbindingFlow to rapidly produce physically plausible dissociation pathways, including novel routes absent from the training data. Crucially, because DD-13M encapsulates the full dynamic progression of unbinding events, it serves as an ideal pre-training resource for kinetics-aware models. When fine-tuned for specific tasks, UnbindingFlow pre-trained on DD-13M demonstrates superior performance in predicting dissociation rates $k_{off}$, effectively bridging the gap between atomic-scale computational sampling and macro-scale pharmacological predictions.

Our scalable framework is distinguished by its highly automated and efficient pipeline, which enables cost-effective expansion of the dataset to enhance predictive accuracy. The method is widely applicable, requiring only an initial bound pose—either experimentally resolved or computationally docked—to initiate trajectory generation, without relying on specific software like DSDP. We have now developed an upgraded pipeline with significantly improved computational efficiency, and a substantially larger dataset is currently in preparation to further increase model generalizability and coverage of diverse biological systems.

This independence from specialized software ensures broad applicability across diverse biological systems. Future developments will focus on integrating advanced sampling methods, such as SinkMeta(65) and Milestoning(66) and higher-precision force fields, to further refine the accuracy of kinetic calculations. By sharing our models and data pipelines, we aim to provide the community with a robust foundation for the next generation of kinetics-aware drug design.

## Acknowledgments

The authors thank Sirui Liu, Xuhan Liu, Zehao Zhou, Cheng Fan, Yize Hao for useful discussion. Computational resources were supported by International Digital Economy Academy and Shenzhen Bay Laboratory supercomputing center. This research was supported by the National Natural Science Foundation of China (22522308 and 22273061 to Y.I.Y.), Shenzhen Hetao Shenzhen-Hong Kong Science and Technology Innovation Cooperation Zone (No. HTHZQSWS-KCCYB-2023052).

## Figures and Tables

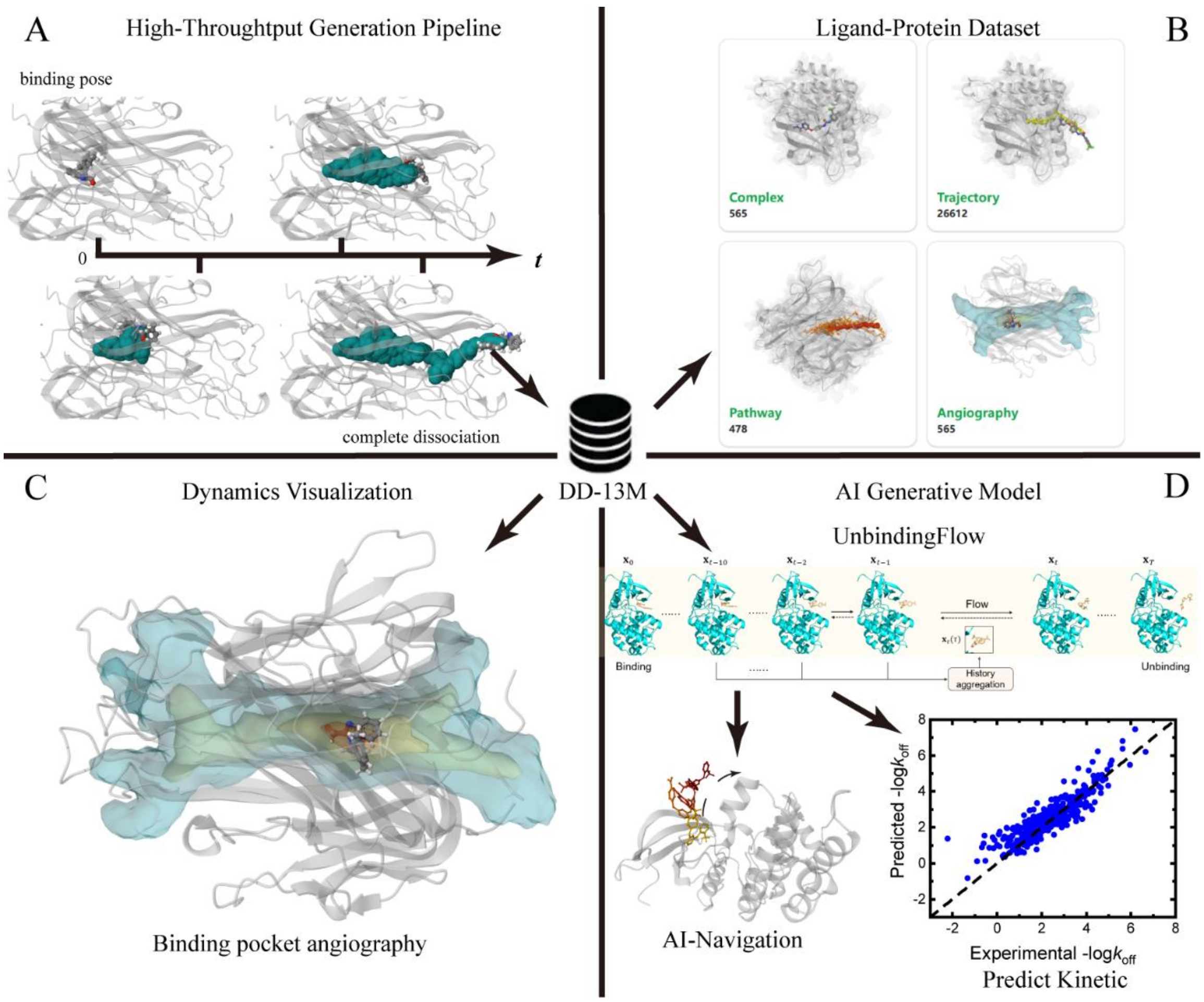

**Figure 1. Overview of DD-13M.** A) High-throughput generation pipeline. Each conformation from the $k_{off}$ dataset underwent a multi-step procedure involving model pre-equilibration, surface definition, and parallel shooting to generate complete dissociation trajectories. B) Distribution statistics of DD-13M. A dedicated web interface is maintained at: https://aimm.szbl.ac.cn/database/ddd/#/home. C) Dynamics visualization. Binding pocket angiography provides a foundational cartographic dataset for studying drug-protein dissociation dynamics. D) AI generative model. We trained a deep equivariant generative model, UnbindingFlow, using the DD-13M dataset to produce dissociation trajectories for novel targets, which statistically significantly improves prediction accuracy for experimental $k_{off}$ rates.

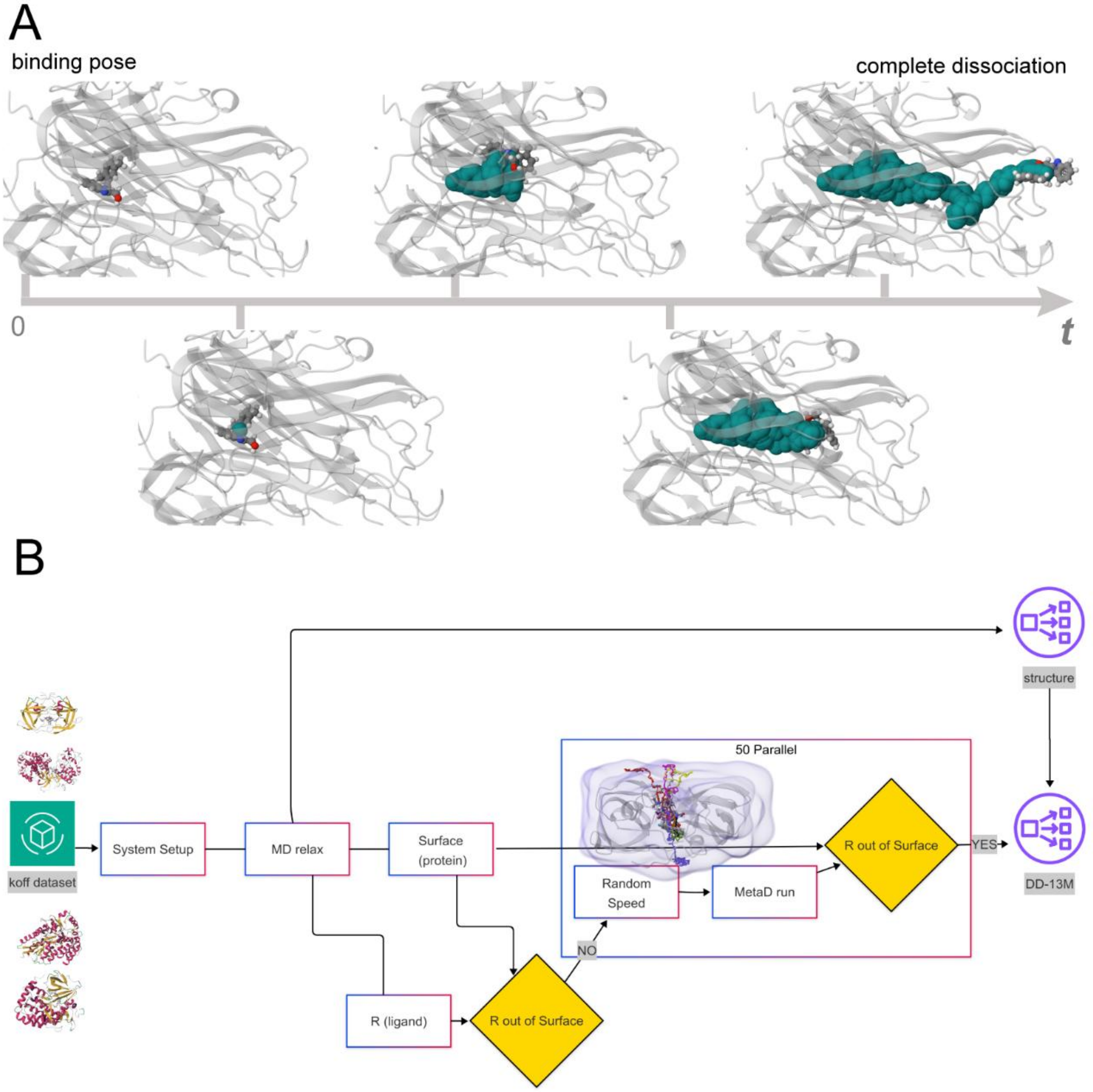

**Figure 2. High-throughput generation pipeline.** A) Snapshots of DD-13M dynamics dataset, taken the drug-protein complex 6OOY as example. The time axis is presented, progressing from left to right. The positions traversed by the centroid coordinates are depicted by filling Gaussian spheres in cyan. B) Each conformation from the $k_{off}$ dataset underwent a multi-step procedure involving model pre-equilibration, surface definition, and parallel shooting to generate complete dissociation trajectories. Details in Supporting Section.

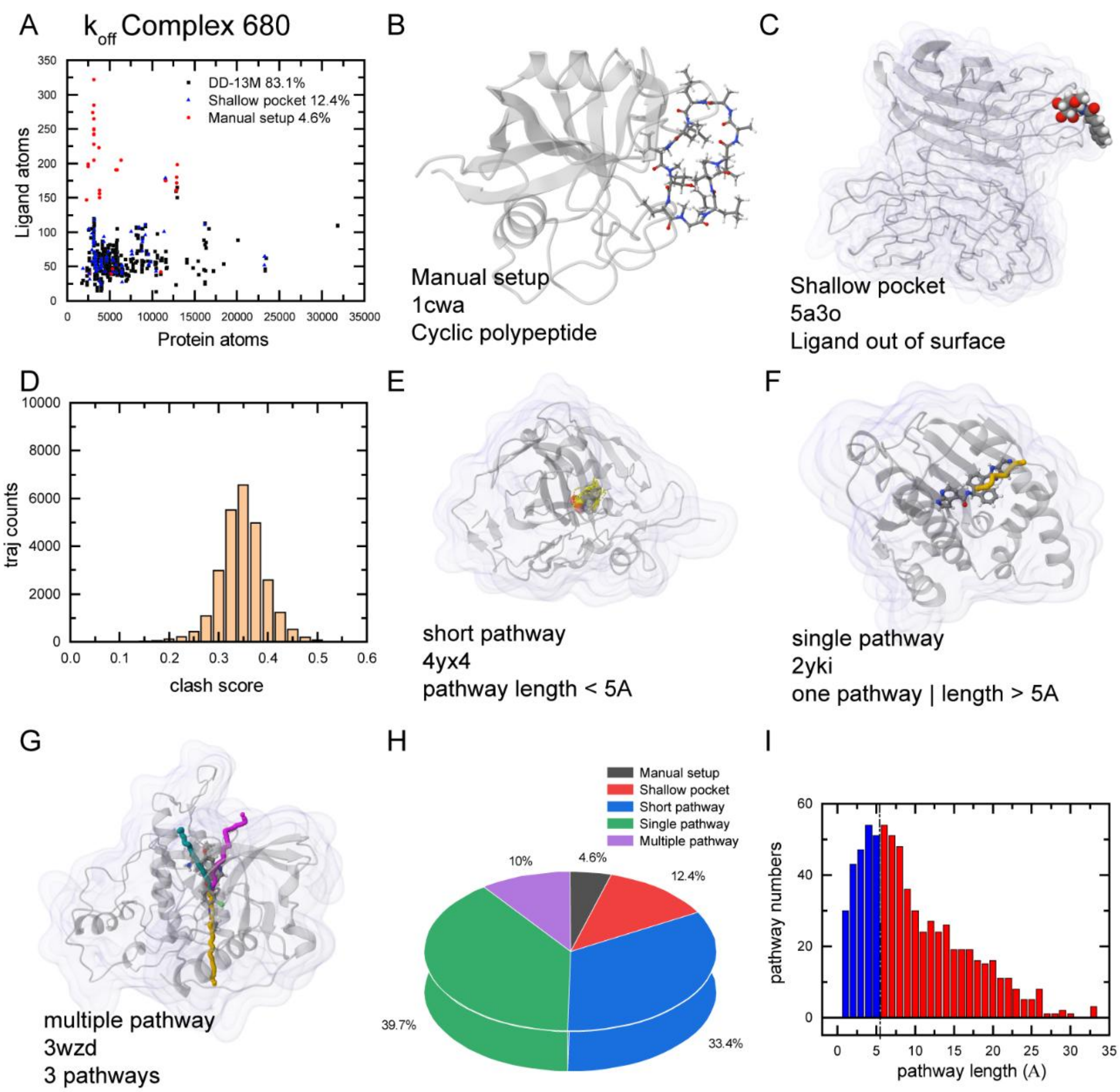

**Figure 3. Database distribution statistics of DD-13M.** A) Sampling results from our modelling approach, where black square represents complexes with trajectory, red circle indicates manual modelling, and blue triangle denote shallow pocket. B) Definition of Manual setup (1cwa): the ligand is a cyclic polypeptide. C) Definition of Shallow pocket (5a3o): in the docking pose, the ligand locates out of the surface. D) Distribution of trajectory clash scores among the 26,612 successful trajectories. E) Definition of short pathway (4yx4): the length of average pathway is shorter than 5 Å. F) Definition of single pathway (2yki): the length of the only pathway is longer than 5 Å. G) Definition of multiple pathway (3wzd): there are more than one pathway. H) Label distribution of $k_{off}$ database: Manual setup:

system which need manual modelling; Shallow pocket: system which the ligand is located outside the protein surface; Short pathway: system with NEB length < 5.0 Å; Single pathway: system with only one pathway; Multiple pathway: system with more than one pathway. I) Clustering distribution of 26,612 trajectories: blue indicates NEB total length < 5.0 Å, which are excluded from the dissociation pathway subset.

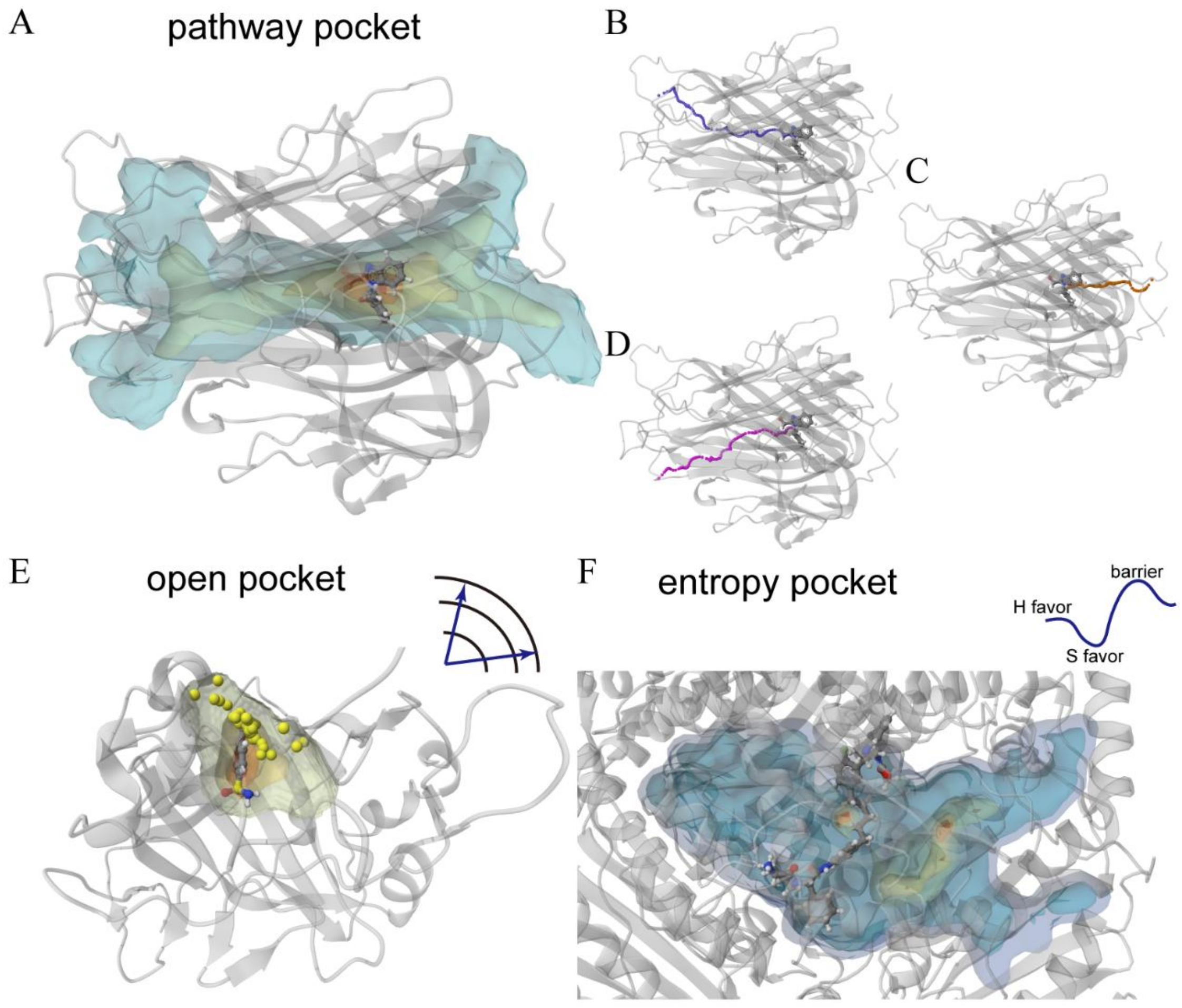

**Figure 4. Binding pocket angiography and dissociation pathways.** A) Binding pocket angiography (BPA), calculated by Eq. (2), details shown in Method. Using the initial protein-ligand binding position as the reference point ($\Delta F = 0$) on the free energy surface, higher $\Delta F$ values correspond to lower ligand binding affinity at that spatial location. The drug-protein complex 6OOY exhibits three major clusters of dissociation trajectories (with trajectory-counts of 5, 37, and 8 respectively). B-D) Three MFEPs using NEB method. E) Open pocket (4yx4): This type of pocket exhibits no dominant pathways, and its properties are primarily determined by the binding affinity strength and the diffusion rate of ligands in solvents, as exemplified in Fig.2C and Fig.2E. F) Entropy dominated pocket (4pfc): this type of pocket, the binding position merely represents the lowest enthalpy point; when entropy contributions are taken into account, positions with even lower free energy exist.

Given that $\Delta F > 0$ at the binding position, the conventional path energy barrier assumption is not suitable for calculations. Such systems constitute less than 5% of the total.

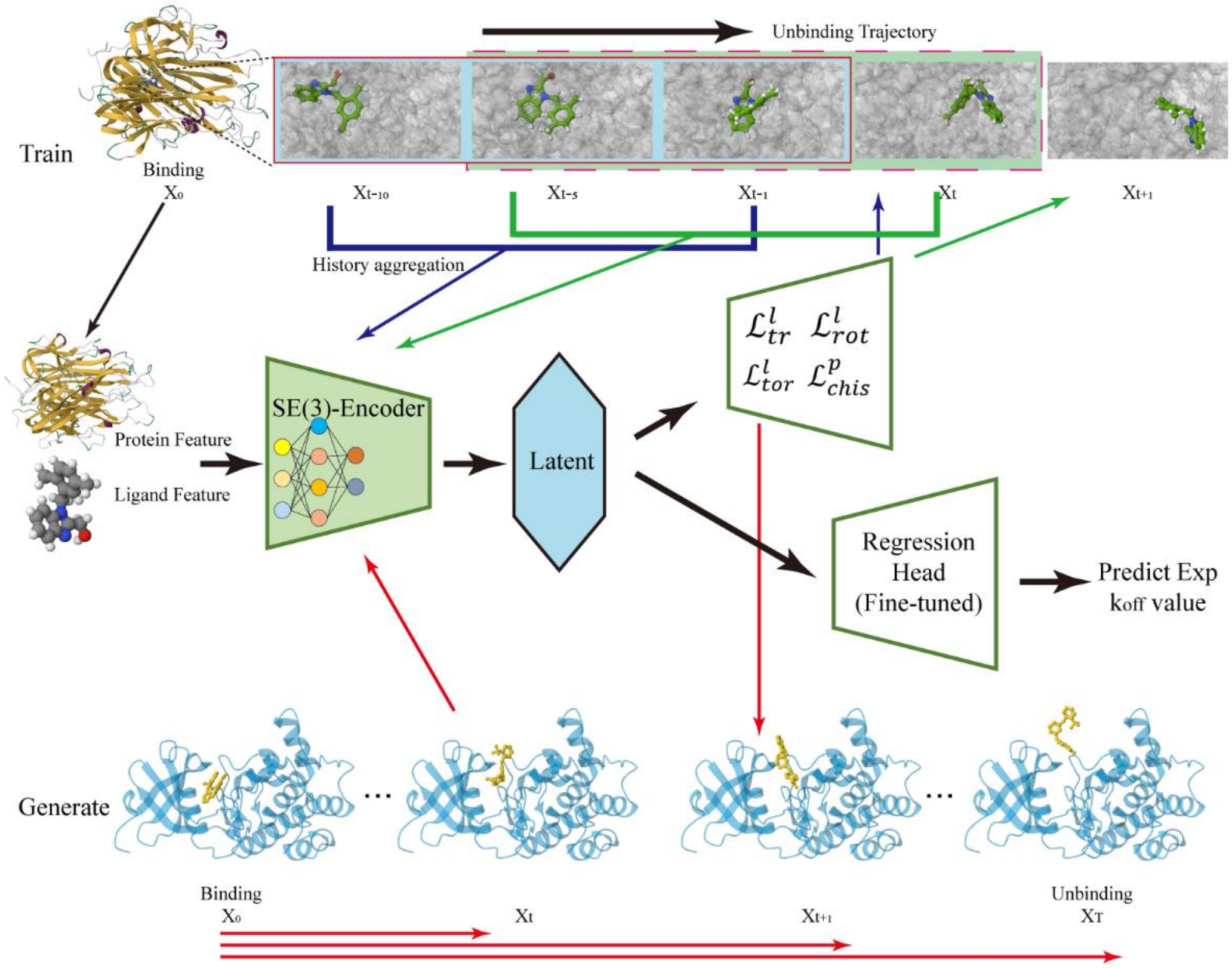

**Figure 5. Overview of AI generative model UnbindingFlow.** The model accepts the drug-protein pair at frame $t-1$ and history ligands at frame as input. The outputs are the predicted updates including the chis angles of each residue, and rotation, translation, torsional angles (also called inner coordinates) of ligand. During training, the model will utilize the whole trajectory from DD13M to calculate the ground truth inner coordinate update between frame *t*-1 to *t*. Implementing these updates to current complex structure gets the next frame structure. We run the model frame-by-frame then we can get the whole unbinding trajectory.

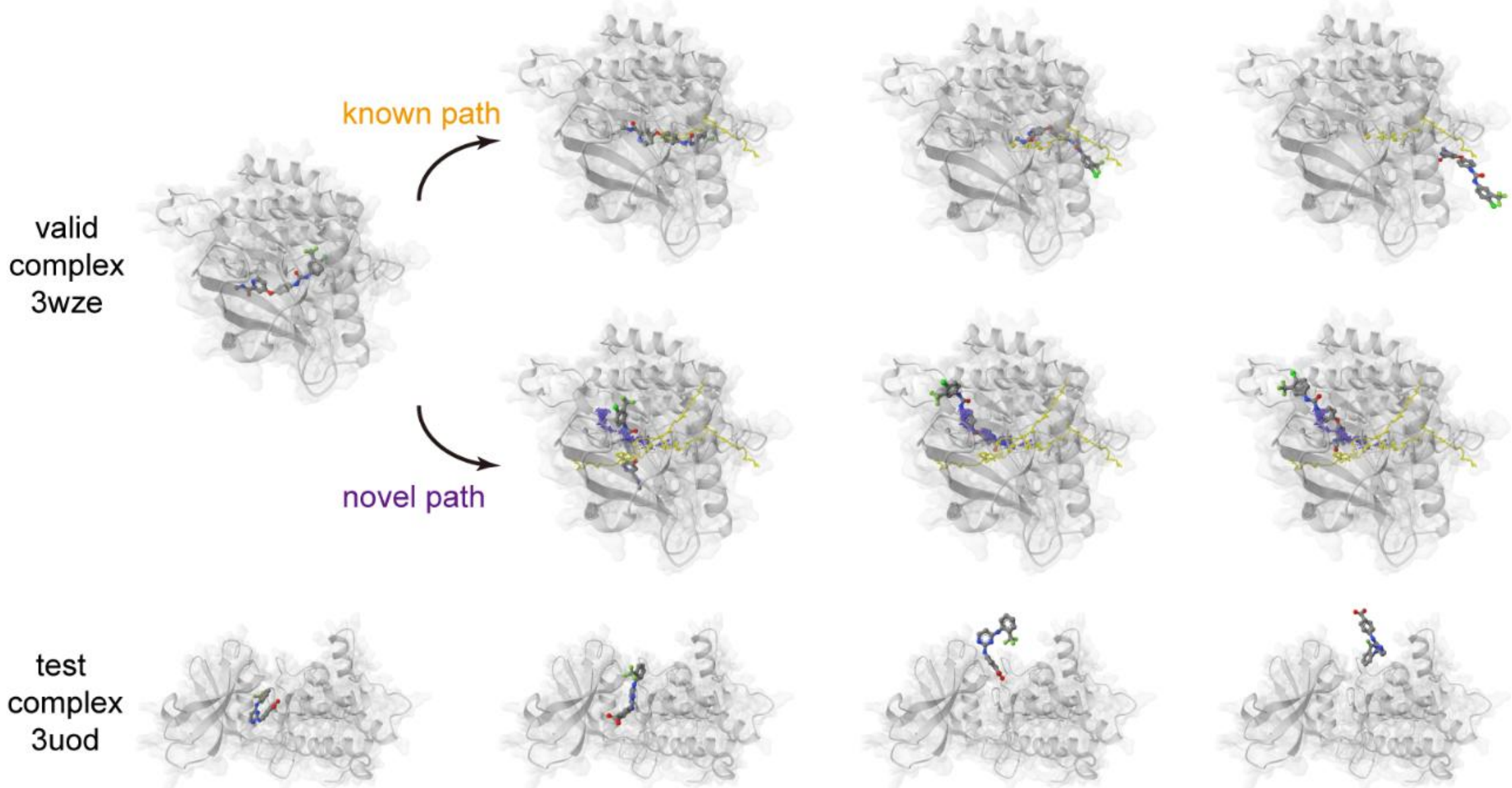

**Figure 6. UnbindingFlow training generative dissociation trajectory visualization, from left to right.** Top: known path from training complex (3wze), the ligand molecule is going outside following the know pathway (yellow balls). Middle: novel path from training complex (3wze), the ligand molecule is going from a novel pathway (purple balls), different from all the three known pathways (yellow balls). Bottom: novel path from novel complex (3uod), which is not included in the DD-13M dataset.

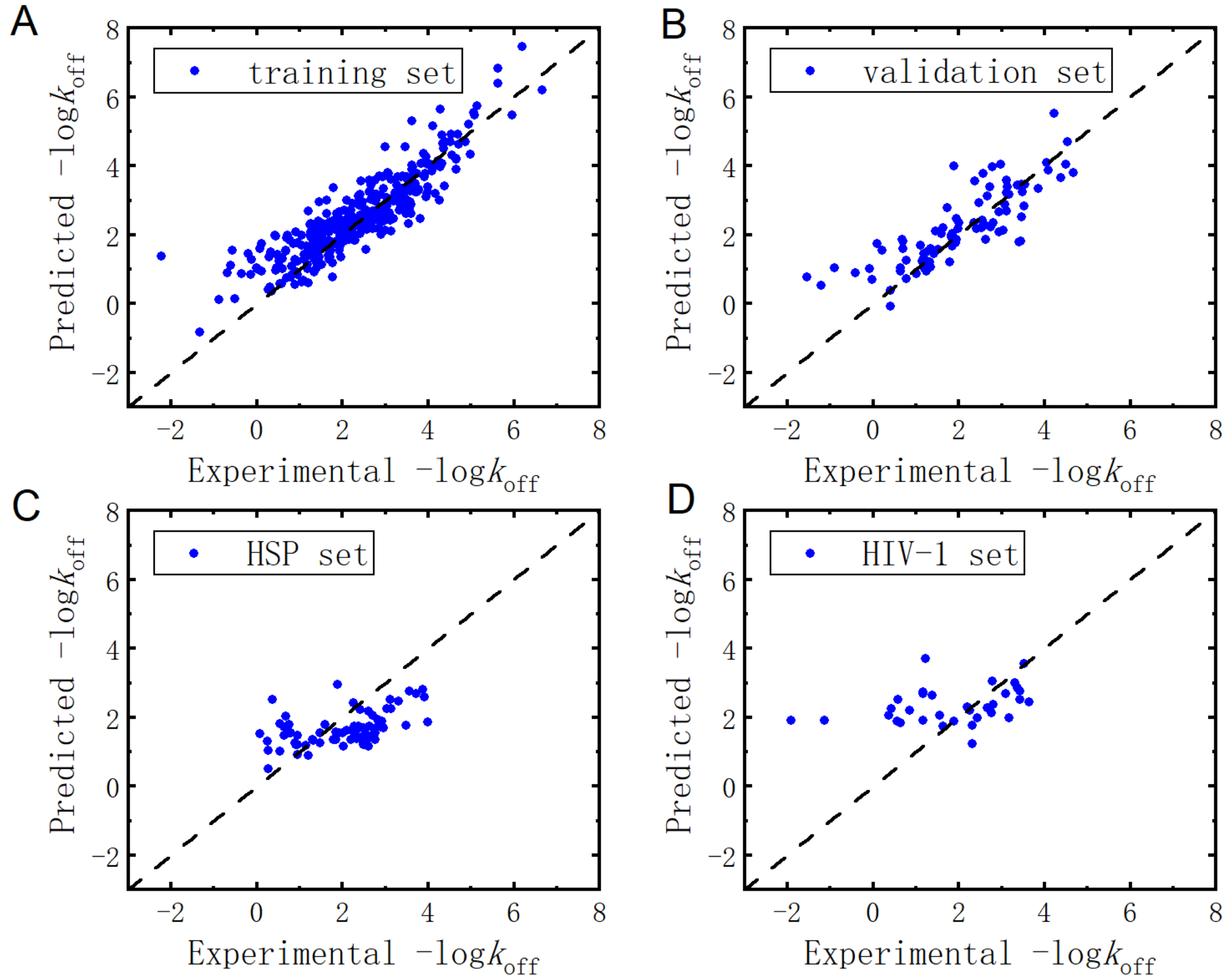

**Figure 7. Correlation between experimentally measured dissociation rate constants (−log $k_{off}$) and the corresponding predicted values for:** A) training set, where $N$ = 332, $R_p$ = 0.8822, and RMSE = 0.6658; B) validation set, where $N$ = 85, $R_p$ = 0.8264, and RMSE = 0.7841; C) test set of HSP90 complexes, where $N$ = 71, $R_p$ = 0.5276, and RMSE = 0.9330; D) test set of HIV-1 protease complexes, where $N$ = 33, $R_p$ = 0.3388, and RMSE = 1.3434.

**Table 1. Performance of Models on $k_{off}$ Prediction.**

| Dataset | | training set | | | validation set | | | HSP set | | | HIV-1 set | | |
|---|---|---|---|---|---|---|---|---|---|---|---|---|---|
| | | $N$ | $R_p$ | RMSE | $N$ | $R_p$ | RMSE | $N$ | $R_p$ | RMSE | $N$ | $R_p$ | RMSE |
| Reported Models | Amangeldiuly et al.[29] | 501 | 0.780 | 0.820 | / | / | / | 100 | **0.560** | 1.160 | 39 | -0.190 | 2.010 |
| | Liu et al.[28] | 407 | 0.968 | 0.474 | 102 | 0.706 | 0.986 | 89 | 0.501 | **0.891** | 39 | 0.306 | 1.602 |
| | Liu et al. (Reproduced) | 332 | **0.969** | 0.452 | 85 | 0.524 | 1.145 | 71 | 0.453 | 0.917 | 33 | 0.264 | 1.498 |
| Unbinding Flow | w/o Pretrain | 332 | 0.223 | 1.335 | 85 | 0.256 | 1.311 | 71 | 0.232 | 0.993 | 33 | 0.056 | 1.393 |
| | +Linear | 332 | 0.962 | **0.444** | 85 | 0.670 | 1.000 | 71 | -0.024 | 1.313 | 33 | -0.180 | 1.459 |
| | +Finetue | 332 | 0.882 | 0.666 | 85 | **0.826** | **0.784** | 71 | 0.528 | 0.933 | 33 | **0.339** | **1.343** |

“UF w/o Pretrain" denotes the UnbindingFlow model without pretraining using DD-13M, trained from scratch solely on PDBbind-koff-2020 static structures with $k_{off}$ labels. "UF+Linear" utilizes the pre-trained UnbindingFlow model with frozen weights, optimizing only the regression head. "UF+Finetune" represents the pre-trained model subjected to full fine-tuning of all parameters. The data for “Liu et al. (Reproduced)” was reproduced based on their published open-source code. While the data for Amangeldiuly et al.’s model was directly adopted from their reported results(29) due to their dataset and code are not open access.